\documentclass[review]{elsarticle}

\usepackage{lineno,hyperref}
\usepackage{amssymb}
\usepackage{graphicx}
\usepackage[ruled,vlined,linesnumbered]{algorithm2e}
\usepackage{float}
\usepackage{color}
\usepackage{soul}
\usepackage{amsmath}
\usepackage{caption}
\usepackage{makecell}
\usepackage{enumitem} 
\usepackage{subcaption}
\modulolinenumbers[5]

\journal{Journal: Internet of Things: Engineering Cyber-Physical Human Systems (Elsevier),}

\begin{document}

\begin{frontmatter}

\title{A Machine Learning based Robust Prediction Model for Real-life Mobile Phone Data}




\author[a,b]{Iqbal H. Sarker$^*$}

\cortext[mycorrespondingauthor]{Corresponding author}
\ead{msarker@swin.edu.au}

\address[a]{Department of Computer Science and Engineering, \\Chittagong University of Engineering and Technology, Bangladesh.}
\address[b]{Department of Computer Science and Software Engineering, \\Swinburne University of Technology, \\ Melbourne, VIC-3122, Australia.}

\begin{abstract}
Real-life mobile phone data may contain \textit{noisy instances}, which is a fundamental issue for building a prediction model with many potential negative consequences. The complexity of the inferred model may increase, may arise over-fitting problem, and thereby the overall \textit{prediction accuracy} of the model may decrease. In this paper, we address these issues and present a \textit{robust prediction model} for real-life mobile phone data of individual users, in order to improve the prediction accuracy of the model. In our robust model, we first effectively \textit{identify} and \textit{eliminate} the noisy instances from the training dataset by determining a \textit{dynamic noise threshold} using \textit{naive Bayes classifier} and \textit{laplace estimator}, which may differ from user-to-user according to their \textit{unique behavioral patterns}. After that, we employ the most popular rule-based machine learning classification technique, i.e., \textit{decision tree}, on the noise-free quality dataset to build the prediction model. Experimental results on the real-life mobile phone datasets (e.g., phone call log) of individual mobile phone users, show the effectiveness of our robust model in terms of precision, recall and f-measure.
\end{abstract}

\begin{keyword}
\texttt{Mobile data mining, machine learning, noise, user behavior modeling, contexts, personalization, prediction model, intelligent systems.}
\end{keyword}

\end{frontmatter}


\section{Introduction}
Nowadays, mobile phone is considered as an essential device of our daily life, as people around the world communicate with each other via mobile phones. According to ITU (International Telecommunication Union), nearly 96.8\% of the world population has been covered by the cellular network coverage, and this coverage even increases up to 100\% of the population in many countries, particularly, in the developed countries like USA, Australia, Canada, UK etc. in the world \cite{number-of-mobile-phone-users}. In another statistics, Sarker et al. \cite{sarker2018MobileDataScience} have shown that users' interest on \textit{``Mobile Phones''} is more and more than other platforms like \textit{``Desktop Computer''} or \textit{``Tablet Computer''} over time. People use mobile phones for various purposes such as voice communication, Internet browsing, apps using, e-mailing, SNS, instant messaging etc. \cite{pejovic2014interruptme}. The smart mobile phones have the ability to log such activities of individual mobile phone users and corresponding contextual information from different sources, such as phone logs \cite{sarker2016phone}, electronic calendars \cite{sarker2016evidence}, or sensors. In this paper, we mainly focus on individuals' diverse \textit{phone call} activities in order to build our \textit{robust prediction model} utilizing their real-life mobile phone data, e.g., phone call log. 

Building an effective prediction model utilizing individual's phone log data is important, as this model can be used to develop various context-aware personalized mobile applications, such as intelligent interruption management system, smart notification management system, intelligent mobile recommender system etc, in order to assist them intelligently in their daily activities. However, the real-life mobile phone data may contain \textit{noisy instances} that may affect on the inferred prediction model, and thereby decrease the effectiveness in terms of prediction accuracy, in these systems. Thus, a \textit{robust prediction model} is a key requirement to build such intelligent systems, which is able to handle noisy instances in mobile phone data and can perform effectively for providing personalized services to assist individuals' in their various day-to-day situations in their real world life.

For the purpose of building a data-driven \textit{robust prediction model} to predict individual's phone call behavior, we use phone log data that consists of individuals' diverse phone call activities, e.g., Accept, Reject, Missed, or making Outgoing call \cite{sarker2018DataCentricSocialContext}, and corresponding contextual information, such as temporal context, spatial context, or social context that have an influence on individuals to make a call handling decision in the real world \cite{sarker2017designing}. For instance, say, on Monday, in the morning a mobile phone user typically rejects (user behavior) the incoming calls if she is in a meeting at her office; however, she accepts (user behavior) if the incoming call is from her elderly mother. Hence, ``Monday morning'' is an example of temporal context, ``at office'' represents user location which is an example of spatial context, and the interpersonal relationship ``mother'', and social situation ``meeting'' are the social contexts relevant to that user. According to \cite{dey2001understanding}, \textit{``context is any information that can be used to characterize the situation of an entity, e.g., an individual mobile phone user"}. Such contextual information may differ from application-to-application according to their relevancy in applications \cite{sarker2018research}. For instance, social relational context might be relevant for one application, say, call interruption management system \cite{sarker2018DataCentricSocialContext}, whereas, it may not be relevant for another type of application, say, mobile notification management system \cite{mehrotra2016prefminer}. 

In the area of mining mobile phone data, both association learning \cite{agrawal1994fast}, and classification learning \cite{quinlan1993} are the most common and popular techniques to build a rule-based user behavior model. However, association learning, e.g., Apriori \cite{agrawal1994fast} produces a large number of redundant rules that makes the prediction model more complex and ineffective \cite{fournier2012mining} \cite{sarker2018mining}. Thus, in this paper, we focus on rule-based classification technique, e.g., decision tree, that can play an important role to build an effective prediction model for individual mobile phone users utilizing their real-life mobile phone data based on multi-dimensional contexts. However, to achieve higher prediction accuracy of the inferred decision-tree based model is challenging. The reason is that classification learning technique requires a quality training data set to build an effective prediction model free from outliers or noise \cite{daza2007algorithm}. Typically, real-world datasets may contain noise or inconsistency instances that may cause over-fitting problem that reduce the prediction accuracy, and consequently, make the prediction model ineffective. Thus, the presence of noise in the dataset is an important issue to effectively modeling mobile user behavior \cite{sarker2018research}. According to \cite{frenay2014classification}, ``noise is anything that obscures the relationship between the features (contexts) in an instance and it's corresponding behavior class''. According to \cite{frenay2014classification}, it is evident that decision trees are badly impacted by noise. Hence, we summarize the effects of noisy instances in the real-life mobile phone data for predicting user behavior as follows:

\begin{itemize}
	\item It may create unnecessary prediction rules that are not interesting to the users and make the rule-set unnecessarily larger.
	\item The complexity of the inferred decision tree based model may increase, due to the number of unnecessary training samples represented as noise.
	\item The presence of noise or inconsistency instances in the dataset may cause over-fitting problem of the decision tree-based model, and thus decrease it's prediction accuracy.
\end{itemize}

Therefore, a \textit{noise elimination} process is required before applying the decision tree classification technique to make the prediction model \textit{robust}, in order to improve the prediction accuracy. According to \cite{zhu2004class}, the performance of a machine learning classification technique depends on two significant factors. First one is the quality of the training data that is used to build the model, and Second one is the competence of the machine learning technique. Therefore, identification and elimination of the noisy instances from a training dataset are required to ensure the quality of the training data in order to build an effective model. 

In this paper, we address the above mentioned issues and present a \textit{robust prediction model} in order to improve the prediction accuracy. In our robust model, we first calculate the conditional probability for all the instances using \textit{naive Bayes classifier} and \textit{laplace estimator}. After that, we dynamically determine a \textit{noise threshold} according to individual's unique behavioral patterns, which may differ from user-to-user. Using this noise threshold, we identify the noisy instances that are unnecessary to build an effective model. Finally, we we employ the prominent rule-based machine learning technique, e.g., decision tree, on the noise-free quality data, to generate a set of prediction rules based on relevant multi-dimensional contexts, for the purpose of building a rule-based effective prediction model.

The contributions are summarized as follows:

\begin{itemize}
	\item We determine a \textit{noise threshold} dynamically by analyzing individual's unique behavioral patterns, for the purpose of identifying inconsistency instances in the dataset.
	
	\item We present a machine learning based \textit{robust prediction model} that effectively predicts individual's phone call behaviors based on multi-dimensional contexts, utilizing their real-life mobile phone data.
	
	\item Our experiments on mobile phone datasets show that this robust model is more effective for predicting user behavior, while comparing with other base models in the area of mining mobile phone data.  
\end{itemize}

 This paper significantly revises and extends our earlier paper \cite{sarker2017improved} in several directions: (i) defining and formulating the problem statement clearly in terms of mathematical notation; (ii) summarizing a number of base modeling approaches that utilize real-life mobile phone data; (iii) designing a decision tree based model and extracting the prediction rules based on multi-dimensional contexts; (iv) conducting additional experiments on real-life mobile phone datasets of individuals; (v) highlighting our key observations in terms of model effectiveness; and finally (vi) summarizing a number of real-world applications of the prediction model for the benefit of end mobile phone users. 
 
The rest of the paper is organized as follows. Section \ref{related work} reviews the related work. We define and formulate the problem statement in Section \ref{Problem-Statement}. We give an overview about the Naive Bayes classifier and Laplace estimator that are used in our model, in Section \ref{Naive Bayes classifier} and Section \ref{Laplacian Estimation} respectively. We present our robust prediction model in section \ref{Our Prediction Model}. We report the experimental results in Section \ref{Experiments}. In section  \ref{Discussion}, we discuss our key observations. We also summarize a number of real life applications of our model in Section \ref{Applications of BOTS}  Finally, Section \ref{Conclusion and Future Work} concludes this paper and highlights the future work.

\section{Related Work}
\label{related work}
Due to the popularity of smart mobile phones with advanced features and context-aware technology, analyzing real-life mobile phone data to model users' behavioral activities, has become an active research area in the current world. A number of researchers have used mobile phone data in order to model and predict user behavior for various purposes. However, these approaches are not \textit{robust} as they do not take into account the issues of noisy or inconsistence instances in the mobile phone data while building the prediction model based on contexts. For instance, Song et al. \cite{song2013exploring} present a study on users' search behavior in order to improve searching relevance based on the log data. Rawassizadeh et al. \cite{rawassizadeh2016scalable} propose a scalable approach utilizing multiple sensor data, for the purpose of mining the daily behavioral patterns of the users. Mukherji et al. \cite{mukherji2014adding} propose an intelligent modeling approach based on multi-dimensional contexts utilizing the mobile phone data. Bayir et al. \cite{bayir2010web} propose an approach for smartphone applications in order to provide the web-based personalized service. In \cite{paireekreng2009time}, Paireekreng et al. have proposed an approach for building personalization mobile game recommendation system. These approaches utilize mobile phone data to build their models based on multi-dimensional contexts. However, they do not take into account the \textit{robustness} to build their models, in which we are interested in.

Besides these approaches, in \cite{mehrotra2016prefminer}, Mehrotra et al. propose a novel interruptibility management system based on mobile phone data. In their approach, they first learns users’ preferences based on multi-contexts about how the users respond with the mobile notifications utilizing mobile phone data. Xu et al. \cite{xu2013preference} have presented a prediction model for the purpose of predicting app usages behavior of mobile phone users. Zhu et al. \cite{zhu2014mining} have presented an approach for the purpose of recommending mobile usages based on relevant contexts utilizing the mobile context log data. In \cite{yu2012towards}, Yu et al. investigate how to exploit user context logs for the purpose of building the personalized context-aware recommendation system through topic models. In \cite{shin2012understanding}, Shin et al. propose an prediction model to predict mobile apps based on mobile phone data.

In addition to these approaches, a number of models based on mobile phone data have been presented for different purposes. For example, an intelligent intrusion administration framework \cite{zulkernain2010mobile}, making application prefetch commonsense on cell phones \cite{parate2013practical}, mining continuous co-event designs on the cell phones \cite{srinivasan2014mobileminer}, mining versatile client propensities \cite{ma2012habit} are displayed dependent on cell phone information. Ozer et al. \cite{ozer2016predicting} propose a way to deal with foresee the area and time of cell phone clients by utilizing consecutive example mining systems. In \cite{do2014and}, Do et al. present a structure for anticipating where clients will go and which application they will use in the following by misusing the rich relevant data from cell phone sensors. In \cite{farrahi2008did}, Farrahi et al. utilize expansive scale cell phone information for examining their day by day schedule practices. In \cite{karatzoglou2012climbing}, Karatzoglou et al. utilize cell phone information in their versatile application suggestion framework. Phithakkitnukoon et al. \cite{phithakkitnukoon2010activity} utilize cell phone information in their investigation to recognize human every day action designs.

In order to predict mobile user navigation patterns, Halvey et al. \cite{halvey2005time} have done their investigation dependent on cell phone information. In their methodology, they fundamentally center around transient setting, as this is a standout amongst the most critical setting sway on client conduct. Phithakkitnukoon et al. \cite{phithakkitnukoon2011behavior} structure a versatile telephone call expectation display using cell phone information. In \cite{jang2015combination}, Jang et al. have done their investigation dependent on cell phone information and demonstrated that clients application uses conduct may differ from client to client after some time. In \cite{henze2011release}, Henze et al. propose a way to deal with locate the best time to convey the portable applications dependent on cell phone information. To identify the appropriate day and age of dynamic applications, Xu et al. \cite{xu2011identifying} propose a methodology by investigating the cell phone information. In \cite{bohmer2011falling}, Bohmer et al. present a way to deal with recognize the pinnacle time of normal application utilizations as per client conduct using their cell phone information.

These approaches discussed above utilize mobile phone data to build their models for different purposes. However, they do not take into account the \textit{robustness}, i.e., issues of inconsistency or noise in the datasets, in their models. Unlike these works, we present a machine learning based \textit{robust prediction model} that eliminates the noisy instances from the real-life mobile phone data, in order to make the model more effective by improving the prediction accuracy.

\section{Definitions and Problem Statement}
\label{Problem-Statement}

This section defines the main notions concerning our \textit{robust prediction model} in order to predict individuals' diverse phone call activities utilizing their mobile phone data. In the following, the notion of mobile phone dataset having phone call activities of individuals with multi-dimensional contexts, is formally stated.

\textbf{Definition 1. (Mobile phone data).} \textit{Let $Con = \{con_1,con_2, ..., con_m\}$  be a set of contexts and $Q = \{q_1,q_2, ..., q_m\}$ the set of corresponding domains. A mobile phone dataset $DS$ is a collection of records, where -}
\textit{
	\begin{enumerate}[label=(\roman*)]
		\item each record $r$ is a set of pairs $(con_i,value_i)$, where $con_i \in Con$, and $value_i \in Q$. For example, if $con_i$ represents the context `spatial', then an example of $value_i$ is `at office'. 
		\item each $con_i \in Con$, also called attribute (context), may occur at most once in any record in DS, and
		\item each record has a particular user activity with mobile phones (e.g., reject phone call).
	\end{enumerate}
}

The mobile phone dataset $DS$ defined above, represents the behavioral data as it contains individual's phone call activities, e.g., rejecting an incoming phone call, in a particular context, e.g., at office.

\textbf{Definition 2. (Phone Call Activity).} \textit{Let, $Act = \{act_1,act_2, ..., act_n\}$  be a set of activity related to the phone calls of an individual user $U$, each action $act_i$ represents a particular activity for that user.} 

In the real world, the common phone call activities of an individual mobile phone user are - (i) answering an incoming phone call, i.e., `Accept', (ii) decline the incoming phone call by the user, i.e., `Reject', (iii) the phone rings but the user misses the call, i.e., Missed, and (iv) making a phone call to a particular person, i.e., `Outgoing' \cite{sarker2018Unavailability}. Each such phone call activity is represented as $act_i$ defined above.

As each activity $act_i$ of an individual user is associated with a particular timestamp (e.g., YYYY-MM-DD hh:mm:ss) recorded by the device, the temporal context can play a primary role to model individual's phone call behavior. Mobile phone records such type of exact temporal information for each activity of an individual, which represents as a time-series that is continuous real-valued. A formal definition of time-series is stated in the following.

\textbf{Definition 3. (Time-Series).} A time-series $T_{series}$ is a sequence of data points ordered in time such that $T_{series} = (t_1, t_2,..., t_m)$, where $t_1, t_2,...,t_m$ are individual observations, each of which contains real-value data and $m$ is the number of observations in a time-series \cite{sarker2017individualized}.

To use time-series data as a temporal context in our robust prediction model, there is a need of time interval, even if only a small interval, e.g., five minutes. The reasons is that exact time is not informative for modeling and predicting user behavior \cite{sarker2017individualized}. Moreover, individuals' day-wise behavior are not identical. For instance, one's phone call behaviors on Monday might be different with her Tuesday's behavior because of her day-to-day situations in the real world. Thus, a day-wise time segment, e.g., Friday[09:00-11:00] that represents the similar behavioral characteristics of an individual mobile phone user in that time period, could be useful. As we aim to build our robust prediction model based on multi-dimensional contexts, in the following, we define the other contexts relevant to our model.

\textbf{Definition 4. (Multi-dimensional Contexts).}  \textit{Let $Con = \{con_1,con_2, ..., con_m\}$  be a set of contexts having influence to make a phone call decision in the real world and $Q = \{q_1,q_2, ..., q_m\}$ the set of corresponding domains related to phone call activity of an individual user $U$. Each context $Con_i$ represents a part of the multi-dimensional contexts.} In addition to the temporal context discussed above, the spatial context can be considered as another dimension of contexts relevant to individual's phone call behavior. For instance, an individual's phone call behavior at her `office' may be well different from her behavior when she is at `home', which represents an example of spatial context. Some examples of these coarse level locations \cite{sarker2014assessing} \cite{eagle2006reality} are office, home, market, store, restaurant, vehicle etc. that can be used to model individual mobile phone users' behavior.

In addition to the above spatio-temporal contexts, social contexts or interpersonal relationship between individuals, mother, might have also an influence on individual mobile phone users to make phone call decisions \cite{sarker2018DataCentricSocialContext}. For example, a user typically `rejects' an incoming phone call during an event official meeting, however, she `answers' if the incoming call comes from her `elderly mother'. Thus, the interpersonal relationship between the caller and the callee has a strong influence to handle phone call decision. 

\textbf{Definition 5. (Interpersonal Relationship).}  \textit{If $U_1$ is a caller and $U_2$ is a callee, then the social relational bonding between $U_1$ and $U_2$ represents the interpersonal relationship in their real life}. 

According to \cite{sarker2018DataCentricSocialContext}, there are several interpersonal relationships, such as mother, family, friend, colleague, boss, significant one, or unknown, can play a strong role to model user phone call behavior of individual mobile phone users in the real world.

\textbf{Definition 6. (Noise).}  \textit{Noise is anything that obscures the relationship between the features or contexts in an instance and it's phone call behavior class. The presence of noisy instances in the training dataset makes the machine learning based behavior prediction model ineffective in terms of prediction accuracy.}

\textbf{Definition 7. (Quality Mobile Phone Data).}  \textit{Let, $N_{total}$ is the total number of instances in the mobile phone data, $N_{noise}$ is the number of noisy instances, then $DS_{quality} = N_{total}$ - $N_{noise}$ represents the quality data that is used to build our robust prediction model for individuals.} \\

\textbf {Problem Statement.} With the above definitions, the main problem we are addressing in this paper is stated as follows:

\textit{Given, a mobile phone dataset $DS$ containing real-life mobile phone data of an individual mobile phone user. Our goal is to identify and eliminate the noisy instances from $DS$ and to model individual's behavior based on the relevant multi-dimensional contexts utilizing the quality data $DS_{quality}$, in order to predict more accurately. In this paper, we present a robust prediction model for the given dataset $DS$, using machine learning techniques, such as naive Bayes classifier, rule-based decision tree classifier, for solving this problem.}

\section{Naive Bayes Classifier}
\label{Naive Bayes classifier}
In our \textit{robust prediction model}, we use Naive Bayes classifier (NBC) to calculate the class conditional probabilities for a given contextual information, in order to identify the noisy instances from the training dataset. NBC in one of the most popular classifiers in the area of data mining and machine learning, which is well-known as a simple probabilistic based method that is able to calculate the class membership probabilities \cite{han2011data}. In NBC, the effect of a contextual attribute in the dataset on a given class is also independent of those of other attributes.

Let $D$ represents a training set of data instances having the contextual attributes and corresponding behavioral class labels. Each instance in the dataset is represented by an n-dimensional context vector, $X = \{x_1, x_2,..., x_n\}$, where n represents the number of attributes of an instance and represented as $\{A_1, A_2,..., A_n\}$. Say, there are $m$ number of behavior classes in the dataset, and represented as $\{C_1, C_2,..., C_m\}$. For a test case of an instance, $X$, the NBC will predict that $X$ belongs to the class with the highest conditional probability, conditioned on $X$. That is, the naive Bayes classifier predicts that the instance $X$ belongs to the class $C_i$, if and only if \cite{han2011data} - \\

$P(C_i|X) > P(C_j|X)$ for $1 \leq j \leq m, j \neq i$ \\

The class $C_i$ for which $P(C_i|X)$ is maximized is called the Maximum Posteriori Hypothesis and defined as \cite{han2011data}:

\begin{equation}
\label{PosterioriHypothesis}
P(C_i|X) =\frac{P(X|C_i)P(C_i)}{P(X)}
\end{equation}

In Bayes theorem shown in Equation \ref{PosterioriHypothesis}, as $P(X)$ is a constant for all classes, only $P(X|C_i)P(C_i)$ needs to be maximized. If the class prior probabilities are not known, then it is commonly assumed that the classes are likely equal, that is, $P(C_1) = P(C_2) = ... = P(C_m)$, and therefore we would maximize $P(X|C_i)$. Otherwise, we maximize $P(X|C_i)P(C_i)$. The class prior probabilities are calculated by $P(C_i) = |C_{i,D}| / |D|$, where $|C_{i,D}|$ is the number of training instances of class $C_i$ in $D$. To compute $P(X|C_i)$ in a dataset with many attributes is extremely computationally expensive. Thus, the naive assumption of class-conditional independence is made in order to reduce computation in evaluating $P(X|C_i)$. This presumes that  the attributes' values are conditionally independent of one another, given the class label of the instance, i.e., there are no dependence relationships among attributes. Thus, Equation \ref{naiveBayesEqu2} and Equation \ref{naiveBayesEqu3} are used to produce $P(X|C_i)$ \cite{han2011data}.

\begin{equation}
\label{naiveBayesEqu2}
P(X|C_i) = \prod_{k=1}^{n} P(x_k|C_i)
\end{equation}

\begin{equation}
\label{naiveBayesEqu3}
P(X|C_i) = P(x_1|C_i) \times P(x_2|C_i) \times ... \times P(x_n|C_i)
\end{equation}

In Equation \ref{naiveBayesEqu2}, $x_k$ refers to the value of attribute $A_k$ for instance $X$. Therefore, these probabilities $P(x_1|C_i), P(x_2|C_i),..., P(x_n|C_i)$ can be easily estimated from the training instances. If the attribute value, $A_k$, is categorical, then $P(x_k|C_i)$ is the number of instances in the class $C_i \in D$ with the value $x_k$ for $A_k$, divided by $|C_{i,D}|$, i.e., the number of instances belonging to the class $C_i \in D$.

To predict the class label of instance $X, P(X|C_i)P(C_i)$ is evaluated for each class $C_i \in D$. The naive Bayes classifier predicts that the class label of instance $X$ is the class $C_i$, if and only if \cite{han2011data} -

$P(X|C_i)P(C_i) > P(X|C_j)P(C_j)$ for $1 \leq j \leq m$ and $j \neq i$

In other words, the predicted class label is the class $C_i$ for which $P(X|C_i)P(C_i)$ is the maximum. NBC has also two main advantages: (a) it requires only one scan of the training data to calculate the probability, and (b) it is easy to use to classify a class according to that probability, which makes the Naive Bayes classifier more popular in the area of classification to predict or classify the class in a given dataset. For instance, for a given particular contexts (office, meeting) in the mobile phone dataset, this classifier is able to predict the behavior (say, reject phone call) by calculating the corresponding conditional probability scanning the relevant contexts over the dataset. However, this classifier is unable to predict the actual behavior class with zero probability \cite{han2011data}. In our robust model, we resolve this issue by using laplace estimator that is discussed in next section.

\section{Laplace Estimator}
\label{Laplacian Estimation}
As in naive Bayes classifier discussed above, we calculate $P(X|C_i)$ as the product of the probabilities $P(x_1|C_i) \times P(x_2|C_i) \times ... \times P(x_n|C_i)$, based on the independence assumption and class conditional probabilities, we will end up with a probability value of zero for some $P(x|C_i)$. This may happen if the attribute value $x$ is never observed in the training data for a particular class $C_i$. Therefore, the generated Equation \ref{naiveBayesEqu3} becomes zeros for such attribute value regardless the values of other attributes. Thus, naive Bayes classifier is unable to predict the class of such test instances. In order to resolve such issues in our model, we use Laplace estimator \cite{cestnik1990estimating} to scale up the values by smoothing factor. In Laplace-estimate, the class probability is defined as \cite{cestnik1990estimating}:

\begin{equation}
\label{laplaceEqu}
P(C = c_i) = \frac{n_c + k}{N + n \times k}
\end{equation}

where $n_c$ is the number of instances satisfying $C = c_i$, $N$ is the number of training instances, $n$ is the number of classes and $k = 1$.

Let's consider a phone call behavior example, for the behavior class `reject' in the training data containing 1000 instances, we have 0 instance with $relationship = unknown$, 990 instances with $relationship = friend$, and 10 instances with $relationship = mother$. The probabilities of these contexts are 0, 0.990 (from 990/1000), and 0.010 (from 10/1000), respectively. On the other hand, according to equation \ref{laplaceEqu}, the probabilities of these contexts would be as follows:

$\frac{1}{1003} = 0.001$,
$\frac{991}{1003} = 0.988$,
$\frac{11}{1003} = 0.011$

\bigskip In this way, we obtain the above non-zero probabilities, rounded up to three decimal places, respectively using Laplace estimator defined above. The ``new'' probability estimates are close to their ``previous'' counterparts, and these values can be used for further processing in our model.

\section{Methodology: Our Robust Prediction Model}
\label{Our Prediction Model}
In this section, we present our \textit{robust prediction model} for real-life mobile phone data of individual users, in order to improve the prediction accuracy of a machine learning based model.

\subsection{Approach Overview}
\begin{figure}[htbp!]
	\centering
	\includegraphics[width=.7\linewidth,keepaspectratio]{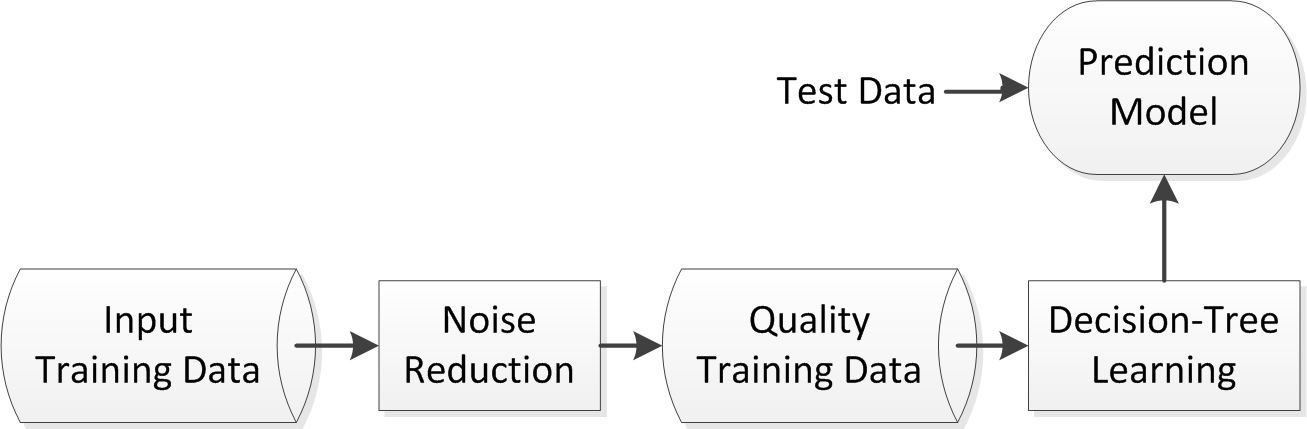}
	\caption{An overview of our robust prediction model}
	\label{fig:block-diagram}
\end{figure}

In our robust model, we first effectively \textit{identify} and \textit{eliminate} the noisy instances from the training dataset by determining a \textit{dynamic noise threshold} that may differ from user-to-user according to their \textit{unique behavioral patterns}. After that, we employ the most popular rule-based machine learning technique, i.e., \textit{decision tree}, on the noise-free quality dataset to build the prediction model. Figure \ref{fig:block-diagram} shows an overview of our robust prediction model that utilizes real-life mobile phone data of individual users.

To achieve our goal, we first use naive Bayes classifier (NBC) discussed above as the basis for noise identification as it calculates the class conditional probability for a given contextual information. Based on such probability values for the contexts and their possible combinations in the datasets, we identify the inconsistency in the dataset. As we aim to build prediction model based on multi-dimensional contexts, such as temporal, spatial, or social context according to the given dataset, the assumption used in NBC for zero probabilities is often unrealistic for predicting individual's phone call behavior \cite{sarker2017aneffectivecall}, which is mentioned earlier. Thus, we further use Laplace-estimator \cite{cestnik1990estimating} discussed above to estimate the conditional probability of contexts. After that, we calculate the ``noise-threshold'' using the generated probability values, in order to identify the noisy instances. As individual's phone call behavioral patterns are not identical in the real world, this noise-threshold may differ according to individual's unique behavioral patterns. 

Once we have identified noisy instances, we eliminate those from the training dataset to get a \textit{noise-free quality data set} that is used to build our robust model. After that, we employ the most popular C4.5 decision tree learning technique \cite{quinlan1993} to generate a set of rules, in order to build our robust prediction model that improves the prediction accuracy while predicting individual's phone call behavior in multi-dimensional contexts.

\subsection{Dynamic Threshold Calculation and Noise Detection}
\label{Our Noise Detection Technique}
In this section, we discuss our \textit{dynamic threshold} based noise detection technique using both the naive Bayes classifier (NBC) and laplace estimator, which may differ from user-to-user according to their behavioral patterns, mentioned earlier. Using NBC, we first calculate the conditional probability for each attribute by scanning the training contextual data. Table \ref{sample-datasets} shows an example of the mobile phone dataset consisting of user phone call behavior, and corresponding multi-dimensional contexts. Each instance contains four attribute or context values (e.g., time, location, situation, and  social relationship between the caller and callee) and corresponding phone call behavior, e.g., Reject or Accept call activities. Based on this information, Table \ref{prior-probability} and Table \ref{conditional-probability} report an example of the prior probabilities for each behavior class and conditional probabilities, respectively. Using these probabilities, we calculate the conditional probability for each given instance. As NBC was implemented under the independence assumption, it estimates zero probabilities if the conditional probability for a single context is zero. In such cases, we apply Laplace-estimator \cite{cestnik1990estimating} to estimate the conditional probability of contexts.

\begin{table*}[htbp!]
	\tiny
	\centering
	\caption{A sample mobile phone dataset}
	\label{sample-datasets}
	\begin{tabular}{|c|c|c|c|c|} 
		\hline
		\textbf{\makecell{DayName \\ TimeSegment}} & \textbf{\makecell{User \\ Location}} & \textbf{\makecell{Social \\ Situation}} & \bf \textbf{\makecell{Social \\ Relationship}} & \textbf{\makecell{Phone Call \\ Behavior}} \\  
		\hline
		Fri[S1] & Office & Meeting & Friend & Reject \\ 
		Fri[S1] & Office & Meeting & Colleague & Reject \\ 
		Fri[S1] & Office & Meeting & Boss & Accept \\ 
		Fri[S1] & Office & Meeting & Friend & Reject \\
		Fri[S2] & Home & Dinner & Friend & Accept \\ 
		Wed[S1] & Office & Seminar & Unknown & Reject \\
		Wed[S1] & Office & Seminar & Colleague & Reject \\
		Wed[S1] & Office & Seminar & Mother & Accept \\
		Wed[S2] & Home & Dinner & Unknown & Accept \\ 
		\hline
	\end{tabular}
\end{table*}

\begin{table}[htbp!]
	\centering
	\tiny
	\caption{An example of prior probabilities}
	\label{prior-probability}
	\begin{tabular}{|c|c|} 
		\hline
		\bf Probability & \bf Value \\  
		\hline
		P(behavior = Reject) & 5/9  \\ 
		P(behavior = Accept) & 4/9  \\ 
		\hline
	\end{tabular}
\end{table}

\begin{table}[h]
	\centering
	\tiny
	\caption{An example of conditional probabilities}
	\label{conditional-probability}
	\begin{tabular}{|c|c|} 
		\hline
		\bf Probability & \bf Value \\  
		\hline
		$P(DayTime = Fri[S1] | behavior = Reject)$ & 3/5  \\ 
		$P(DayTime = Fri[S1] | behavior = Accept)$ & 1/4  \\ 
		$P(DayTime = Fri[S2] | behavior = Reject)$ & 0/5  \\ 
		$P(DayTime = Fri[S2] | behavior = Accept)$ & 1/4  \\ 
		$P(DayTime = Wed[S1] | behavior = Reject)$ & 2/5  \\ 
		$P(DayTime = Wed[S1] | behavior = Accept)$ & 1/4  \\ 
		$P(DayTime = Wed[S2] | behavior = Reject)$ & 0/5  \\ 
		$P(DayTime = Wed[S2] | behavior = Accept)$ & 1/4  \\ 
		$P(Location = Office | behavior = Reject)$ & 5/5  \\ 
		$P(Location = Office | behavior = Accept)$ & 2/4  \\ 
		$P(Location = Home | behavior = Reject)$ & 0/5  \\ 
		$P(Location = Home | behavior = Accept)$ & 2/4  \\ 
		$P(Situation = Meeting | behavior = Reject)$ & 3/5  \\ 
		$P(Situation = Meeting | behavior = Accept)$ & 1/4  \\ 
		$P(Situation = Seminar | behavior = Reject)$ & 2/5  \\ 
		$P(Situation = Seminar | behavior = Accept)$ & 1/4  \\ 
		$P(Situation = Dinner | behavior = Reject)$ & 0/5  \\ 
		$P(Situation = Dinner | behavior = Accept)$ & 2/4  \\ 
		$P(Relationship = Friend | behavior = Reject)$ & 2/5  \\ 
		$P(Relationship = Friend  | behavior = Accept)$ & 1/4  \\ 
		$P(Relationship = Colleague | behavior = Reject)$ & 2/5  \\ 
		$P(Relationship = Colleague  | behavior = Accept)$ & 0/4  \\ 
		$P(Relationship = Boss | behavior = Reject)$ & 0/5  \\ 
		$P(Relationship = Boss  | behavior = Accept)$ & 1/4  \\ 
		$P(Relationship = Mother | behavior = Reject)$ & 0/5  \\ 
		$P(Relationship = Mother  | behavior = Accept)$ & 1/4  \\ 
		$P(Relationship = Unknown | behavior = Reject)$ & 1/5  \\ 
		$P(Relationship = Unknown  | behavior = Accept)$ & 1/4  \\ 
		\hline
	\end{tabular}
\end{table}

\begin{figure}[htbp!]
	\centering
	\includegraphics[width=.7\linewidth,keepaspectratio]{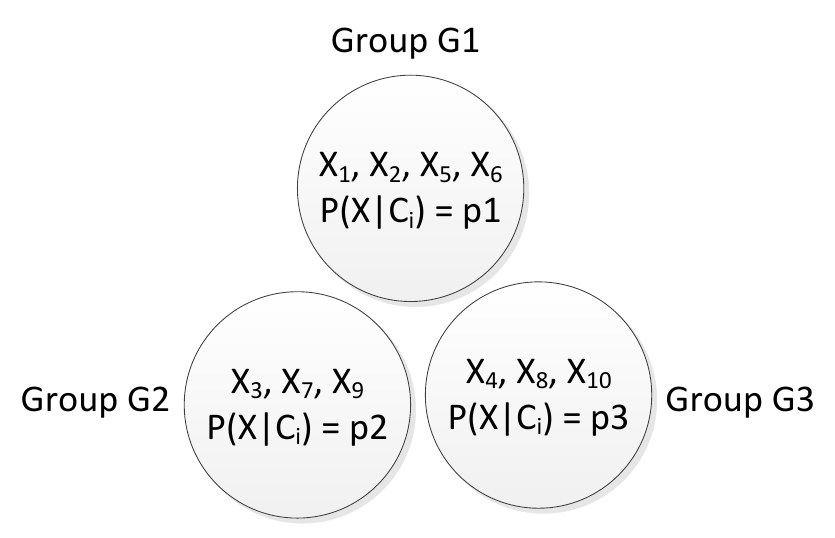}
	\caption{An example of instances-group based on probability}
	\label{fig:group}
\end{figure}

Once we have calculated conditional probability for each instance, we differentiate between the purely classified instances and misclassified instances using these values. ``Purely classified'' instances are those for which the predicted class and the original class is same. If different class is found then the instances are considered as ``misclassified'' \cite{farid2014hybrid}. After that, we generate the instance groups by taking into account all the distinct probabilities as separate group values. Figure \ref{fig:group} shows an example of instance groups $G1, G2, G3$ for the instances ${X_1,X_2,...,X_{10}}$, where $G1$ consists of 5 instances with probability $p1$, $G2$ consists of 3 instances with probability $p2$, and finally $G3$ consists of 3 instances with probability $p3$. We then identify the group among the purely classified instances for which the probability is minimum. This minimum probability is considered as ``noise-threshold''. Finally, the instances in misclassified list differentiated earlier, for those the probability does not satisfy the noise threshold, are identified as noise.

\begin{algorithm}[h]
	\tiny
	\caption{Noise Detection Technique}
	\label{alg:noise-detection}
	\SetKwInOut{Data}{Data}
	\SetAlgoLined
	\Data{Training dataset: $D = {X_1,X_2,...,X_n}$ // Training dataset, $D$, which contains a set of training instances and their associated class labels.}
	\KwResult{noise list: $noise_{list}$}
	
	\BlankLine
	
	\ForEach{class, $C_i \in D$}
	{  
		Find the prior probabilities, $P(C_i)$.	
	}
	\ForEach{attribute value, $A_{i} \in D$}
	{  
		Find the class conditional probabilities, $P(A_{i}|C_i)$.	
	}
	\ForEach{training instance, $X_i \in D$}
	{  
		Find the conditional probability, $P(X_i|C_i)$ \\
		
		\If{$P(X_i|C_i)$ $==$ 0}
		{
			//apply Laplace Estimator\\
			recalculate the conditional probability, $P(X_i|C_i)$ using Laplace Estimator
		}
		\If{$X_i$ is misclassified}
		{
			$misClass_{list} \leftarrow X_i$ \\
			// store the probabilities for all misclassified instances. \\
			$misPro_{list} \leftarrow P(X_i|C_i)$	  
		}
		\Else
		{
			$pureClass_{list} \leftarrow X_i$ \\
			// store the probabilities for all purely classified instances.\\
			$purePro_{list} \leftarrow P(X_i|C_i)$   
		}
	}
	
	$T_{noise} $=$ findMIN(purePro_{list})$ // noise threshold	
	
	\ForEach{instance, $x_i \in misClass_{list}$}
	{  
		Find the conditional probability, $P(X_i|C_i)$ from $misPro_{list}$
		
		\If{$P(X_i|C_i)$ $<$ $T_{noise}$} 
		{
			// store instances as noise. \\
			$noise_{list} \leftarrow X_i$ 
		}
	}
	return $noise_{list}$
	
\end{algorithm}

The process for identifying noise is set out in Algorithm \ref{alg:noise-detection}. Input data includes training dataset: $D = {X_1,X_2,...,X_n}$, which contains a set of training instances and their associated class labels and output data is the list of noisy instances. For each class, we calculate the prior probabilities $P(C_i)$. After that for each attribute value, we calculate the class conditional probabilities $P(A_{i}|C_i)$. For each training instance, we calculate the conditional probabilities $P(X_i|C_i)$. We then check whether it is non-zero. If we get zero probabilities, we then recalculate the conditional probabilities $P(X_i|C_i)$ using Laplace Estimator. Based on these probability values, we then check whether the instances are misclassified or purely classified and store all misclassified instances $misClass_{list}$ with corresponding probabilities in $misPro_{list}$. Similarly, we also store all purely classified instances $pureClass_{list}$ with corresponding probabilities in $purePro_{list}$. We then identify the minimum probability from $purePro_{list}$ as noise threshold. Finally, this algorithm returns a set of noisy instances $noise_{list}$ for a particular dataset.

Rather than arbitrarily determine the threshold, our algorithm dynamically identifies the noise threshold according to individual's behavioral patterns and identify noisy instances based on this threshold. As individual's phone call behavioral patterns are not identical in the real world, this noise-threshold for identifying noisy instances changes dynamically according to individual's unique behavioral patterns.  

\subsection{Decision Tree based Model and Rule Generation}
In this section, we discuss our decision-tree based prediction model for generating a set of rules. It is developed based on a basic C4.5 algorithm \cite{quinlan1993}. Given a training dataset,  $DS = \{X_1, X_2,..., X_m\}$, where $m$ represents the sample size. Each instance is represented by an n-dimensional attribute vector, $X = \{x_1, x_2,..., x_n\}$, depicting $n$ measurements made on the instance from $n$ attributes, respectively, $\{A_1, A_2,..., A_n\}$. The training data also belong to a set of classes $C = {C_1,C_2,...,C_m}$. A decision tree is a rule-based classification tree associated with $DS$, which is a structure that includes a root node, internal nodes, leaf nodes, and their associate arcs. The topmost node in the tree is called the root. Each internal node denotes a test on a context attribute, $A_i$, (e.g., social relationship), and each leaf node denotes the behavioral outcome class, $C_i$, of that test which is represented by a behavior class label (e.g., reject phone call). Each arc is associated with a particular context value, e.g., friend as a social relationship context. After building the complete decision tree, rules are extracted from the tree for the purpose of using in prediction for a particular test case. 

\begin{figure}[htbp!]
	\centering
	\includegraphics[width=.8\textwidth]{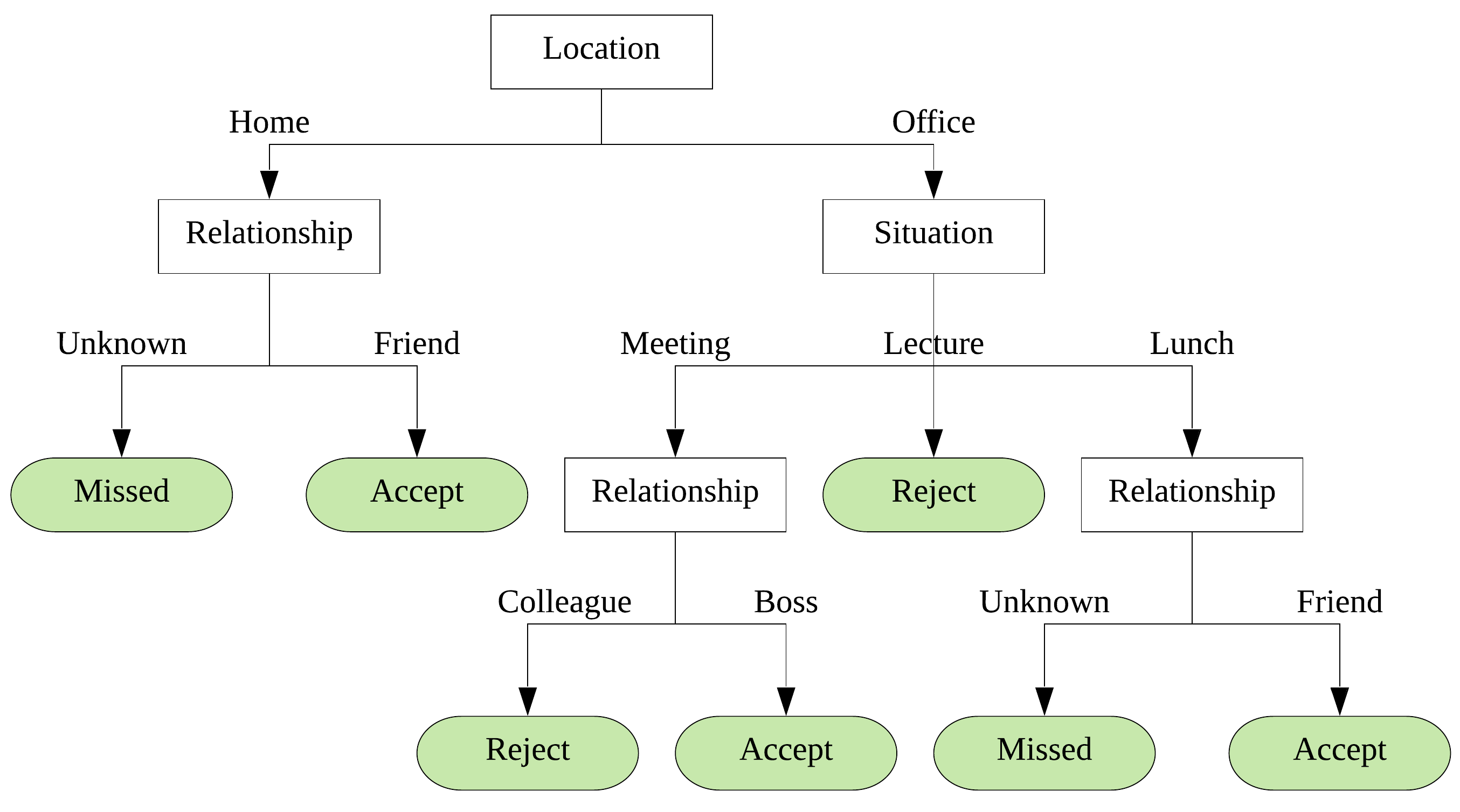}
	\caption{An example of a decision-tree based model.}
	\label{fig:tree}
\end{figure} 

Figure \ref{fig:tree} shows an example of such a decision tree for multi-dimensional contexts, where locational context values are $\{home, office\}$, interpersonal social relational context values are $\{boss, friend, colleague, unknown\}$, social situation values are $\{meeting, lecture, lunch\}$, and user phone call behavior classes are $\{Accept, Reject, Missed\}$. Once the tree has been generated, rules are extracted by traversing the tree from root node to each leaf node. The followings are the examples of the produced rules $\{R_1, R_2,...,R_7\}$ generated from the tree, shown in Figure \ref{fig:tree}.

\noindent 
$R_1: {Home, Unknown \Rightarrow Missed}$ \\
$R_2: {Home, Friend \Rightarrow Accept}$ \\
$R_3: {Office, Meeting, Colleague \Rightarrow Reject}$ \\
$R_4: {Office, Meeting, Boss \Rightarrow Accept}$ \\
$R_5: {Office, Lecture \Rightarrow Reject}$ \\
$R_6: {Office, Lunch, Unknown \Rightarrow Missed}$ \\
$R_7: {Office, Lunch, Friend \Rightarrow Accept}$ \\

Rule $R_1$ states that the user misses the incoming calls from unknown people, when she is at home. However, accepts the incoming calls at home if the calls come from her friends, which is stated in Rule $R_2$. Similarly,  Rule $R_3$ states that the user rejects the incoming calls of her colleagues, when she is in a meeting at her office. However, she accepts the incoming calls from her boss, which is stated in Rule $R_4$. Similarly, the other rules $R_5, R_6, R_7$ state the behaviors of the user in other relevant contexts. These generated rules based on relevant multi-dimensional contexts are able to predict individual's phone call behavior, and can be used to build a context-aware intelligent system, to assist them in their daily activities.

\section{Experiments and Evaluation}
\label{Experiments}
In this section, we describe our experimental results for the real-life mobile phone datasets of individual users. We also present an experimental evaluation comparing our robust prediction model with the existing approaches for predicting user phone call behavior.

\subsection{Real-life Mobile Phone Dataset}
We have conducted experiments on ten phone log datasets collected by Massachusetts Institute of Technology (MIT) for their Reality Mining project~\cite{eagle2006reality}. Each dataset represents the phone call activities of an individual mobile phone user. These individuals are faculty, staff, and students. The datasets include people with different types of calling patterns and call distributions. These datasets contain multi-dimensional contexts, such as temporal, location, and social context, and corresponding phone call activities of individuals. Table \ref{Datasets descriptions} describes each call log dataset represented as $(DS01, DS02, ..., DS10)$ for ten individual mobile phone users respectively.

\begin{table}[htbp!]
	\centering
	\tiny
	\caption{Datasets descriptions of individual mobile phone users.}
	\label{Datasets descriptions}
	\begin{tabular}{|c|c|c|} 
		\hline
		\bf \makecell{Mobile Phone Dataset \\ (individual users)} & \bf \makecell{Data Size \\ (instances)} & \bf \makecell{Time Period of \\ User Activities} \\  
		\hline
		DS01 & 9204 & 9 months \\ 
		\hline
		DS02 & 5121 & 4 months \\ 
		\hline
		DS03 & 506 & 2 months \\ 
		\hline
		DS04 & 8518 & 9 months \\ 
		\hline
		DS05 & 2982 & 6 months \\ 
		\hline
		DS06 & 7040 & 6 months \\
		\hline
		DS07 & 1578 & 3 months \\
		\hline
		DS08 & 8606 & 4 months \\
		\hline
		DS09 & 7752 & 9 months \\
		\hline
		DS10 & 3798 & 9 months \\
		\hline
	\end{tabular}
\end{table}

\subsection{Data-Preprocessing}
Hence, we pre-process the raw data available in the dataset. As seen in the dataset, the mobile phone records both accepting and rejecting calls as incoming calls, we distinguish accept and reject calls by using the call duration. According to \cite{sarker2016behavior}, if the call duration is greater than 0 then the call has been accepted; if it is equal to 0 then the call has been rejected. Overall, we take into account four types of phone call behavioral classes, such as accept/answer call, reject/decline call, missed call, and making an outgoing call, of all the individuals. In Table \ref{tab:context-examples-call}, we summarize all these activities (behavior classes) and corresponding contextual information that are taken into account in our experiments. In order to pre-process the temporal data for building this model, we use our earlier bottom-up behavior-oriented time segmentation (BOTS) technique \cite{sarker2017individualized} that extracts individual's behavior oriented time segments according to the similar behavioral characteristics. For instance, Friday[09:00-11:00], Monday[12:00-13:00], Saturday[15:30-18:45] are the examples of the generated segments using this technique. We also generate data-centric social relational context using mobile phone data, where ``each unique mobile phone number represents a particular one-to-one relationship'' \cite{sarker2018DataCentricSocialContext}. For instance, mother's phone number (047XXXX231) represents one relation ($Rel_{1}$), while friend's phone number (047XXXX232) represents another relation ($Rel_{2}$) etc.

\begin{table*}[htbp!]
	\begin{center}
		\caption{Examples of various phone call activities and corresponding contextual information}
		
		\label{tab:context-examples-call}
		\begin{tabular}{l|c} 
			\textbf{Context Category} & \textbf{Context Examples}\\
			\hline
			Temporal Context & \makecell{User's activity occuring date (YYYY-MM-DD), \\ time (hh:mm:ss), period (e.g., 1 hour, 10:00am-12:00pm), \\ weekday (e.g., Monday), weekend (e.g., Saturday), etc.}\\
			\hline
			
			Spatial Context & \makecell{User's coarse level location such as \\ at office, work, home, market, on the way, restaurant, \\ vehicle, playground etc.}\\
			\hline
			
			Social Context & \makecell{User's social relationship between individuals \\ such as mother, friend, family, colleague,  boss, \\ significant one, unknown, etc.}\\
			\hline
			
			Call Activities & \makecell{Answering an incoming call, Decline an incoming call, \\ Missed call, and making an Outgoing call}\\
			\hline
		\end{tabular}
	\end{center}
\end{table*}

\subsection{Evaluation Metric}
In order to measure the prediction accuracy, we compare the predicted response with the actual response (i.e., the ground truth) and compute the accuracy in terms of:

\begin{itemize}
	\item Precision: ratio between the number of phone call behaviors that are correctly predicted and the total number of behaviors that are predicted (both correctly and incorrectly). If TP and FP denote true positives and false positives then the formal definition of precision is \cite{han2011data}:
	
	\begin{equation}
	Precision = \frac{TP}{TP + FP}
	\end{equation}
	
	\item Recall: ratio between the number of phone call behaviors that are correctly predicted and the total number of behaviors that are relevant. If TP and FN denote true positives and false negatives then the formal definition of recall is \cite{han2011data}:
	
	\begin{equation}
	Recall = \frac{TP}{TP + FN}
	\end{equation}
	
	\item F-measure: a measure that combines precision and recall is the harmonic mean of precision and recall. The formal definition of F-measure is \cite{han2011data}:
	
	\begin{equation}
	Fmeasure = 2 * \frac{Precision * Recall}{Precision + Recall}
	\end{equation}
	
\end{itemize}

\subsection{Evaluation Results}
To evaluate our model, we employ the most popular cross validation technique, N-fold \cite{han2011data}, in machine learning, where we use N=10 to measure the outcome. The 10-fold cross validation breaks data into 10 sets. It trains the prediction model on 9 sets and tests it's performance using the remaining one set. This repeats 10 times and we take a mean accuracy rate.

\begin{table}[htbp!]
	\centering
	\tiny
	\caption{An example of noisy instances identified from the training dataset for a single iteration of N-fold.}
	\label{Noisy-instances}
	\begin{tabular}{|c|c|c|} 
		\hline
		\bf \makecell{Mobile Phone Dataset \\ (individual users)} & \bf \makecell{Noisy \\ Instances} \\  
		\hline
		DS01 & 5.23\% \\ 
		\hline
		DS02 & 3.52\% \\ 
		\hline
		DS03 & 2.03\% \\ 
		\hline
		DS04 & 4.12\% \\ 
		\hline
		DS05 & 0.24\% \\ 
		\hline
		DS06 & 4.89\% \\
		\hline
		DS07 & 3.49\% \\
		\hline
		DS08 & 4.70\% \\
		\hline
		DS09 & 5.35\% \\
		\hline
		DS10 & 4.67\% \\
		\hline
	\end{tabular}
\end{table}

\begin{figure}[htbp!]
	\centering
	\includegraphics[width=.6\linewidth,keepaspectratio]{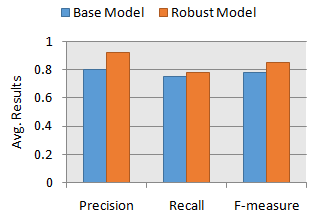}
	\caption{Effectiveness comparison results}
	\label{fig:comparison}
\end{figure}

To show the effectiveness of our robust model, we compare the prediction results of existing approaches with our model, in terms of precision, recall and f-measure, defined earlier. In our experiment, existing approaches are those that use real-life mobile phone data for modeling without taking into account the \textit{robustness}, i.e, the quality of the training data for modeling. For the purpose of effectiveness comparison with our robust model, we represent such type of existing works as `Base Model', shown in Figure \ref{fig:comparison}. In order to make fare comparison of both models, we use the same datasets described above, in our experiments. Figure \ref{fig:comparison} shows the overall impact of noisy instances in mobile phone data on the effectiveness of the prediction model. In Table \ref{Noisy-instances}, we also show the number of noisy instances identified by our approach for an iteration, which have an impact on the overall prediction accuracy. For another iteration, the number of noisy instances may change. The reason is that, we use 10-fold cross validation technique for evaluation, where 90\% data is used as training data to build the model in each iteration. In another iteration, the training data is changed according to the procedure of N-fold cross validation, and consequently, the number of noisy instances may also change.

To show the overall effectiveness in terms of prediction accuracy of our robust model, Figure \ref{fig:comparison} shows the relative comparison of precision, recall and f-measure, by calculating the average results for all the datasets described earlier. If we observe Figure \ref{fig:comparison}, we find that our robust model consistently outperforms base model for predicting individual's phone call behavior in terms of precision, recall and f-measure. The main reason is that existing base models do not take into account the robustness while predicting user behavior, and the resulting accuracy is consequently low. On the other-hand, our robust model effectively handles the noisy instances by analyzing the behavioral patterns of individual mobile phone users, and thus improves the prediction accuracy in relevant contexts.

\section{Discussion}
\label{Discussion}
Overall, our robust prediction model is completely individualized and behavior-oriented. Contrasted with the existing applicable methodologies, the user behavior prediction accuracy in terms of precision, recall and f-measure, is improved when our robust prediction model is applied, as shown in Figure \ref{fig:comparison}. Although, the robust model is a two-step process, it is effective while demonstrating user behavior using real life cell phone information. The following are few key discoveries from our study. 

\begin{itemize}
	\item Real-world datasets may contain noisy instances. Such noisy instances may diminish the precision of expectations, increment the multifaceted nature of model induction process because of noisy training samples. To recognize such cases, determining a noise threshold according to individual's unique behavioral patterns is the key term in our methodology. However, the threshold can differ from user-to-user, as the behavioral patterns of individual mobile phone users are not identical in the real world.
	
	\item Machine learning based approaches, such as using Naive Bayes classifier and Laplace Estimator can play an important role for identifying and eliminating noisy instances from the training dataset, in order to prepare the quality training data for an effective modeling. Such techniques also can help to calculate the dynamic noise-threshold by analyzing the given contextual data, in order to identify the noisy instances.
	
	\item We have observed a significantly lower prediction accuracy when using the base model compared to our robust model. The reason is that existing base models do not take into account the impact of noisy instances in their models. Consequently, the prediction results found using base models have low precision, recall and f-measure, shown in Figure \ref{fig:comparison}.
	
	\item Our approach does not depend on any particular number of contexts. However, we take into account temporal, spatial, and social contexts, as these are relevant to our problem domain. For another problem domain in mobile phones, the corresponding relevant contexts can be used in our approach.	
\end{itemize}

\section{Real-Life Applications}
\label{Applications of BOTS}
As our model is able to predict future behavior of individual mobile phone users based on relevant contextual information utilizing their mobile phone data, this model can be used in various real-life applications to assist them intelligently. Hence, we summarize a number of relevant applications. These are:

\begin{itemize}
	\item \textit{Intelligent Voice Communication:} Managing incoming call interruptions can be an important application of this model. For this, phone call log data having different types of behaviors with incoming phone calls in different contexts, is needed to build the model. Such application will be helpful for individual mobile phone users to assist them in their daily activities, as mobile phones are considered to be `always on, always connected' device in the real world, but the users are not always attentive and responsive to incoming phone calls, because of their various day-to-day situations \cite{chang2015investigating}. Such call intrusions may make humiliating circumstance in an official situation, e.g., meeting, as well as influence in different exercises like inspecting patients by a specialist or driving a vehicle and so on. These call interferences may likewise diminish laborer execution, expanded blunders and worry in a workplace \cite{pejovic2014interruptme}. Therefore, in order to minimize such interruptions, a robust model utilizing individual's phone call response behavioral data collected from related sources having relevant contexts such as temporal, spatial, social or interpersonal relationship between caller and callee, can be used to build an intelligent call interruption management system. Similarly, building robust model utilizing outgoing call related data can be used to build a smart call reminder system that can help to keenly looks through the attractive contact from the substantial contact list and reminds the user to make a phone call to a specific individual in a specific settings.
	
	\item \textit{Mobile Notification Management:} Notifications are a core feature of mobile phones \cite{sahami2014large}. In the real world, a variety of smart mobile applications use notifications in order to inform the the users about various kinds of events, such as the arrival of a message, a new comment on their posts on social networks, like Facebook, LinkedIn, Twitter, the updating availability of an application like WhatsApp, Skype, Viber, various news or just to send them reminders or alerts \cite{sahami2014large}. However, many of them (mobile phone notifications) are neither useful nor relevant to the interests of the users. As a result, such useless notifications are considered disruptive and potentially annoying to the users \cite{mehrotra2016prefminer} \cite{iqbal2010notifications}. Individual's behavioral rules based on user's contextual information, might be able to manage such notifications intelligently. For examples, one individual always dismisses promotional email notifications; she accepts birthday reminder notifications of Facebook mostly at night, when she is at home; does not accept Viber or Whatsapp notifications from unknown persons at office. Therefore, building robust model utilizing individual's notification handling behavioral data collected from related sources having relevant contexts such as temporal, spatial, notification type, can be used to build an intelligent notification management system.
	
	\item \textit{Mobile Recommendation System:} Mobile recommender system is one of the most important applications of user behavior model, as it helps the user to find the most satisfying service by reducing search effort and information overload. For instance, a particular mobile app recommendation among a huge number of installed apps (e.g., Multimedia, Facebook, Gmail, Youtube, Skype, Game, Microsoft outlook, etc.) according to users current contextual information (temporal, spatial or others), and preference could be useful for the users. For this relevant mobile phone data having various contexts and corresponding app usages can be used for building the app recommendation model for the purpose of identifying which app is preferred by a particular user under a certain context to provide personalized context-aware recommendation of different mobile phone apps for the mobile phone users. Similarly, other recommendation services can be generated from relevant mobile phone data using our robust model in order to get better prediction accuracy in various contexts.
\end{itemize}

\section{Conclusion and Future Work}
\label{Conclusion and Future Work}
In this paper, we have presented a \textit{robust prediction model} for the real-life mobile phone data, in order to improve the prediction accuracy. In our robust model, we have effectively handled the noisy instances from the training dataset by determining a dynamic noise threshold using naive Bayes classifier and laplace estimator, which may differ from user-to-user according to their unique behavioral patterns. In order to build the complete model, we have also employed the most popular rule-based machine learning technique, i.e., decision tree, on the noise-free quality dataset. Experimental results on multi-contextual phone call log datasets indicate that compare to the existing approaches, our robust model improves the prediction accuracy in terms of precision, recall and f-measure.

In future work, we plan to use this model for building real-world mobile applications discussed above, to assess the effectiveness of this model in application level.

\section*{Acknowledgment}
The author would like to thank Prof. Jun Han, Swinburne University of Technology, Australia, Dr. Alan Colman, Swinburne University of Technology, Australia, and Dr. Ashad Kabir, Charles Sturt University, Australia for their relevant discussions.

\bibliographystyle{plain}
\bibliography{bibfile/noise-bibfile}

\begin{thebibliography}{10}

\bibitem{agrawal1994fast}
Rakesh Agrawal and Ramakrishnan Srikant.
\newblock Fast algorithms for mining association rules.
\newblock In {\em Proceedings of the International Joint Conference on Very
  Large Data Bases, Santiago Chile, pp.$\sim$487--499.}, volume 1215, 1994.

\bibitem{bayir2010web}
Murat~Ali Bayir, Murat Demirbas, and Ahmet Cosar.
\newblock A web-based personalized mobility service for smartphone
  applications.
\newblock {\em The Computer Journal}, 54(5):800--814, 2010.

\bibitem{bohmer2011falling}
Matthias B{\"o}hmer, Brent Hecht, Johannes Sch{\"o}ning, Antonio Kr{\"u}ger,
  and Gernot Bauer.
\newblock Falling asleep with angry birds, facebook and kindle: a large scale
  study on mobile application usage.
\newblock In {\em Proceedings of the International Conference on Human computer
  interaction with mobile devices and services, Stockholm, Sweden, 30 August -
  2 September, pp.$\sim$47--56. ACM, New York, USA}, 2011.

\bibitem{cestnik1990estimating}
Bojan Cestnik et~al.
\newblock Estimating probabilities: a crucial task in machine learning.
\newblock In {\em ECAI}, volume~90, pages 147--149, 1990.

\bibitem{chang2015investigating}
Yung-Ju Chang and John~C Tang.
\newblock Investigating mobile users' ringer mode usage and attentiveness and
  responsiveness to communication.
\newblock In {\em Proceedings of the International Conference on Human-Computer
  Interaction with Mobile Devices and Services, Copenhagen, Denmark, 24-27
  August, pp.$\sim$6--15. ACM, New York, USA}, 2015.

\bibitem{daza2007algorithm}
Luis Daza and Edgar Acuna.
\newblock An algorithm for detecting noise on supervised classification.
\newblock In {\em Proceedings of WCECS-07, the 1st World Conference on
  Engineering and Computer Science}, pages 701--706, 2007.

\bibitem{dey2001understanding}
Anind~K Dey.
\newblock Understanding and using context.
\newblock {\em Personal and ubiquitous computing}, 5(1):4--7, 2001.

\bibitem{do2014and}
Trinh Minh~Tri Do and Daniel Gatica-Perez.
\newblock Where and what: Using smartphones to predict next locations and
  applications in daily life.
\newblock {\em Pervasive and Mobile Computing}, 12:79--91, 2014.

\bibitem{eagle2006reality}
Nathan Eagle and Alex~Sandy Pentland.
\newblock Reality mining: sensing complex social systems.
\newblock {\em Personal and ubiquitous computing}, 10(4):255--268, 2006.

\bibitem{farid2014hybrid}
Dewan~Md Farid, Li~Zhang, Chowdhury~Mofizur Rahman, M~Alamgir Hossain, and
  Rebecca Strachan.
\newblock Hybrid decision tree and na{\"\i}ve bayes classifiers for multi-class
  classification tasks.
\newblock {\em Expert Systems with Applications}, 41(4):1937--1946, 2014.

\bibitem{farrahi2008did}
Katayoun Farrahi and Daniel Gatica-Perez.
\newblock What did you do today?: discovering daily routines from large-scale
  mobile data.
\newblock In {\em Proceedings of the International Conference on Multimedia,
  Vancouver, British Columbia, Canada, 26-31 October, pp.$\sim$849--852. ACM,
  New York, USA}, 2008.

\bibitem{fournier2012mining}
Philippe Fournier-Viger and Vincent~S Tseng.
\newblock Mining top-k non-redundant association rules.
\newblock In {\em International Symposium on Methodologies for Intelligent
  Systems}, pages 31--40. Springer, 2012.

\bibitem{frenay2014classification}
Beno{\^\i}t Fr{\'e}nay and Michel Verleysen.
\newblock Classification in the presence of label noise: a survey.
\newblock {\em IEEE transactions on neural networks and learning systems},
  25(5):845--869, 2014.

\bibitem{halvey2005time}
Martin Halvey, Mark~T Keane, and Barry Smyth.
\newblock Time based segmentation of log data for user navigation prediction in
  personalization.
\newblock In {\em Proceedings of the International Conference on Web
  Intelligence, Compiegne, France, 19-22 September, pp.$\sim$636--640. IEEE
  Computer Society, Washington, DC, USA.}, 2005.

\bibitem{han2011data}
Jiawei Han, Jian Pei, and Micheline Kamber.
\newblock {\em Data mining: concepts and techniques}.
\newblock Elsevier, Amsterdam, Netherlands, 2011.

\bibitem{henze2011release}
Niels Henze and Susanne Boll.
\newblock Release your app on sunday eve: finding the best time to deploy apps.
\newblock In {\em Proceedings of the International Conference on Human Computer
  Interaction with Mobile Devices and Services, Stockholm, Sweden, 30 August -
  2 September, pp.$\sim$581--586. ACM, New York, USA}, 2011.

\bibitem{iqbal2010notifications}
Shamsi~T Iqbal and Eric Horvitz.
\newblock Notifications and awareness: a field study of alert usage and
  preferences.
\newblock In {\em Proceedings of the 2010 ACM conference on Computer supported
  cooperative work}, pages 27--30. ACM, 2010.

\bibitem{jang2015combination}
Bo-Ram Jang, Yunseok Noh, Sang-Jo Lee, and Seong-Bae Park.
\newblock A combination of temporal and general preferences for app
  recommendation.
\newblock In {\em Proceedings of the International Conference on Big Data and
  Smart Computing(BigComp), Jeju, South Korea, 9-11 February,
  pp.$\sim$178--185. IEEE Computer Society, Washington, DC, USA}, 2015.

\bibitem{karatzoglou2012climbing}
Alexandros Karatzoglou, Linas Baltrunas, Karen Church, and Matthias B{\"o}hmer.
\newblock Climbing the app wall: enabling mobile app discovery through
  context-aware recommendations.
\newblock In {\em Proceedings of the International Conference on Information
  and knowledge management, Maui, Hawaii, USA, 29 October - 02 November,
  pp.$\sim$2527--2530. ACM, New York, USA}, 2012.

\bibitem{ma2012habit}
Haiping Ma, Huanhuan Cao, Qiang Yang, Enhong Chen, and Jilei Tian.
\newblock A habit mining approach for discovering similar mobile users.
\newblock In {\em Proceedings of the International Conference on World Wide
  Web, Lyon, France, 16-20 April, pp.$\sim$231--240. ACM, New York, USA}, 2012.

\bibitem{mehrotra2016prefminer}
Abhinav Mehrotra, Robert Hendley, and Mirco Musolesi.
\newblock Prefminer: mining user's preferences for intelligent mobile
  notification management.
\newblock In {\em Proceedings of the International Joint Conference on
  Pervasive and Ubiquitous Computing, Heidelberg, Germany, 12-16 September,
  pp.$\sim$1223--1234. ACM, New York, USA.}, 2016.

\bibitem{mukherji2014adding}
Abhishek Mukherji and Vijay Srinivasan.
\newblock Adding intelligence to your mobile device via on-device sequential
  pattern mining.
\newblock In {\em Proceedings of the International Joint Conference on
  Pervasive and Ubiquitous Computing, Seattle, WA, USA, 13-17 September,
  pp.$\sim$1005-1014. ACM, New York, USA.}, 2014.

\bibitem{ozer2016predicting}
Mert Ozer, Ilkcan Keles, Hakki Toroslu, Pinar Karagoz, and Hasan Davulcu.
\newblock Predicting the location and time of mobile phone users by using
  sequential pattern mining techniques.
\newblock {\em The Computer Journal}, 59(6):908--922, 2016.

\bibitem{paireekreng2009time}
Worapat Paireekreng, Kowit Rapeepisarn, and Kok~Wai Wong.
\newblock Time-based personalised mobile game downloading.
\newblock In {\em Transactions on Edutainment II, pp.$\sim$59--69}. 2009.

\bibitem{parate2013practical}
Abhinav Parate, Matthias B{\"o}hmer, David Chu, Deepak Ganesan, and Benjamin~M
  Marlin.
\newblock Practical prediction and prefetch for faster access to applications
  on mobile phones.
\newblock In {\em Proceedings of the International Joint Conference on
  Pervasive and Ubiquitous Computing, Zurich, Switzerland, 8-12 September,
  pp.$\sim$275--284. ACM, New York, USA}, 2013.

\bibitem{pejovic2014interruptme}
Veljko Pejovic and Mirco Musolesi.
\newblock Interruptme: designing intelligent prompting mechanisms for pervasive
  applications.
\newblock In {\em Proceedings of the International Joint Conference on
  Pervasive and Ubiquitous Computing, Seattle, WA, USA, 13-17 September,
  pp.$\sim$897--908. ACM, New York, USA.}, 2014.

\bibitem{phithakkitnukoon2011behavior}
Santi Phithakkitnukoon, Ram Dantu, Rob Claxton, and Nathan Eagle.
\newblock Behavior-based adaptive call predictor.
\newblock {\em ACM Transactions on Autonomous and Adaptive Systems},
  6(3):21:1--21:28), 2011.

\bibitem{phithakkitnukoon2010activity}
Santi Phithakkitnukoon and Teerayut Horanont.
\newblock Activity-aware map: Identifying human daily activity pattern using
  mobile phone data.
\newblock In {\em In Salah A.A., Gevers T., Sebe N., Vinciarelli A. (eds) Human
  Behavior Understanding. Lecture Notes in Computer Science, Springer, Berlin,
  Heidelberg}. 2010.

\bibitem{quinlan1993}
J.~Ross Quinlan.
\newblock C4.5: Programs for machine learning.
\newblock {\em Machine Learning}, 1993.

\bibitem{rawassizadeh2016scalable}
Reza Rawassizadeh, Elaheh Momeni, Chelsea Dobbins, Joobin Gharibshah, and
  Michael Pazzani.
\newblock Scalable daily human behavioral pattern mining from multivariate
  temporal data.
\newblock {\em IEEE Transactions on Knowledge and Data Engineering},
  28(11):3098--3112, 2016.

\bibitem{sahami2014large}
Alireza Sahami~Shirazi, Niels Henze, Tilman Dingler, Martin Pielot, Dominik
  Weber, and Albrecht Schmidt.
\newblock Large-scale assessment of mobile notifications.
\newblock In {\em Proceedings of the SIGCHI Conference on Human Factors in
  Computing Systems}, pages 3055--3064. ACM, 2014.

\bibitem{sarker2014assessing}
Hillol Sarker, Moushumi Sharmin, Amin~Ahsan Ali, Md~Mahbubur Rahman, Rummana
  Bari, Syed~Monowar Hossain, and Santosh Kumar.
\newblock Assessing the availability of users to engage in just-in-time
  intervention in the natural environment.
\newblock In {\em Proceedings of the 2014 ACM International Joint Conference on
  Pervasive and Ubiquitous Computing}, pages 909--920. ACM, 2014.

\bibitem{sarker2018MobileDataScience}
Iqbal~H Sarker.
\newblock Mobile data science: Towards understanding data-driven intelligent
  mobile applications.
\newblock {\em EAI Endorsed Transactions on Scalable Information Systems},
  2018.

\bibitem{sarker2018research}
Iqbal~H Sarker.
\newblock Research issues in mining user behavioral rules for context-aware
  intelligent mobile applications.
\newblock {\em Iran Journal of Computer Science, Springer}, pages 1--11, 2018.

\bibitem{sarker2018Unavailability}
Iqbal~H Sarker.
\newblock Silentphone: Inferring user unavailability based opportune moments to
  minimize call interruptions.
\newblock {\em EAI Endorsed Transactions on Mobile Communications and
  Applications}, 2018.

\bibitem{sarker2018DataCentricSocialContext}
Iqbal~H Sarker.
\newblock Understanding the role of data-centric social context in personalized
  mobile applications.
\newblock {\em EAI Endorsed Transactions on Context-aware Systems and
  Applications}, 2018.

\bibitem{sarker2016behavior}
Iqbal~H Sarker, Alan Colman, Muhammad~Ashad Kabir, and Jun Han.
\newblock Behavior-oriented time segmentation for mining individualized rules
  of mobile phone users.
\newblock In {\em Proceedings of the 2016 IEEE International Conference on Data
  Science and Advanced Analytics (IEEE DSAA), Montreal, Canada.}, pages
  488--497. IEEE, 2016.

\bibitem{sarker2016phone}
Iqbal~H Sarker, Alan Colman, Muhammad~Ashad Kabir, and Jun Han.
\newblock Phone call log as a context source to modeling individual user
  behavior.
\newblock In {\em Proceedings of the 2016 ACM International Joint Conference on
  Pervasive and Ubiquitous Computing (Ubicomp): Adjunct, Germany}, pages
  630--634. ACM, 2016.

\bibitem{sarker2017individualized}
Iqbal~H Sarker, Alan Colman, Muhammad~Ashad Kabir, and Jun Han.
\newblock Individualized time-series segmentation for mining mobile phone user
  behavior.
\newblock {\em The Computer Journal, Oxford University, UK}, 61(3):349--368,
  2018.

\bibitem{sarker2016evidence}
Iqbal~H Sarker, Muhammad~Ashad Kabir, Alan Colman, and Jun Han.
\newblock Evidence-based behavioral model for calendar schedules of individual
  mobile phone users.
\newblock In {\em Proceedings of the 2016 IEEE International Conference on Data
  Science and Advanced Analytics (IEEE DSAA), Montreal, Canada.}, pages
  584--593. IEEE, 2016.

\bibitem{sarker2017designing}
Iqbal~H Sarker, Muhammad~Ashad Kabir, Alan Colman, and Jun Han.
\newblock Designing architecture of a rule-based system for managing phone call
  interruptions.
\newblock In {\em Proceedings of the 2017 ACM International Joint Conference on
  Pervasive and Ubiquitous Computing and Proceedings of the 2017 ACM
  International Symposium on Wearable Computers, USA}, pages 898--903. ACM,
  2017.

\bibitem{sarker2017aneffectivecall}
Iqbal~H Sarker, Muhammad~Ashad Kabir, Alan Colman, and Jun Han.
\newblock An effective call prediction model based on noisy mobile phone data.
\newblock In {\em Proceedings of the 2017 ACM International Joint Conference on
  Pervasive and Ubiquitous Computing: Adjunct}. ACM, 2017.

\bibitem{sarker2017improved}
Iqbal~H Sarker, Muhammad~Ashad Kabir, Alan Colman, and Jun Han.
\newblock An improved naive bayes classifier-based noise detection technique
  for classifying user phone call behavior.
\newblock In {\em Proceedings of the 2017 Australian Data Mining Conference
  (AusDM 2017), Melbourne, Australia}. Springer, 2017.

\bibitem{sarker2018mining}
Iqbal~H Sarker and Flora~D Salim.
\newblock Mining user behavioral rules from smartphone data through association
  analysis.
\newblock In {\em Proceedings of the 22nd Pacific-Asia Conference on Knowledge
  Discovery and Data Mining (PAKDD), Melbourne, Australia}, pages 450--461.
  Springer, 2018.

\bibitem{shin2012understanding}
Choonsung Shin, Jin-Hyuk Hong, and Anind~K Dey.
\newblock Understanding and prediction of mobile application usage for smart
  phones.
\newblock In {\em Proceedings of the International Conference on Ubiquitous
  Computing, Pittsburgh, Pennsylvani, 5-8 September, pp.$\sim$173--182. ACM,
  New York, USA}, 2012.

\bibitem{song2013exploring}
Yang Song, Hao Ma, Hongning Wang, and Kuansan Wang.
\newblock Exploring and exploiting user search behavior on mobile and tablet
  devices to improve search relevance.
\newblock In {\em Proceedings of the International Conference on World Wide
  Web, Rio de Janeiro, Brazil, 13-17 May, pp.$\sim$1201-1212. ACM, New York,
  USA}, 2013.

\bibitem{srinivasan2014mobileminer}
Vijay Srinivasan, Saeed Moghaddam, and Abhishek Mukherji.
\newblock Mobileminer: Mining your frequent patterns on your phone.
\newblock In {\em Proceedings of the International Joint Conference on
  Pervasive and Ubiquitous Computing, Seattle, WA, USA, 13-17 September,
  pp.$\sim$389--400. ACM, New York, USA.}, 2014.

\bibitem{number-of-mobile-phone-users}
International~Telecommunication Union.
\newblock Measuring the information society.
\newblock In {\em Technical report, http://www.itu.int/en/ITU-D/Statistics/
  Documents/publications/misr2015/MISR2015-w5.pdf}, 2015.

\bibitem{xu2011identifying}
Qiang Xu, Jeffrey Erman, Alexandre Gerber, Zhuoqing Mao, Jeffrey Pang, and
  Shobha Venkataraman.
\newblock Identifying diverse usage behaviors of smartphone apps.
\newblock In {\em Proceedings of the ACM SIGCOMM conference on Internet
  measurement conference, Berlin, Germany, 2-4 November, pp.$\sim$329--344.
  ACM, New York, USA}, 2011.

\bibitem{xu2013preference}
Ye~Xu, Mu~Lin, Hong Lu, Giuseppe Cardone, Nicholas Lane, Zhenyu Chen, Andrew
  Campbell, and Tanzeem Choudhury.
\newblock Preference, context and communities: a multi-faceted approach to
  predicting smartphone app usage patterns.
\newblock In {\em Proceedings of the International Symposium on Wearable
  Computers, Zurich, Switzerland, 8-12 September, pp.$\sim$69--76. ACM, New
  York, USA}, 2013.

\bibitem{yu2012towards}
Kuifei Yu, Baoxian Zhang, Hengshu Zhu, Huanhuan Cao, and Jilei Tian.
\newblock Towards personalized context-aware recommendation by mining context
  logs through topic models.
\newblock In {\em Proceedings of the Pacific-Asia Conference on Knowledge
  Discovery and Data Mining, Kuala Lumpur, Malaysia, May 29 - June 01,
  pp.$\sim$431--443. Springer-Verlag Berlin, Heidelberg}, 2012.

\bibitem{zhu2014mining}
Hengshu Zhu, Enhong Chen, Hui Xiong, Kuifei Yu, Huanhuan Cao, and Jilei Tian.
\newblock Mining mobile user preferences for personalized context-aware
  recommendation.
\newblock {\em ACM Transactions on Intelligent Systems and Technology (TIST)},
  5(4):58, 2014.

\bibitem{zhu2004class}
Xingquan Zhu and Xindong Wu.
\newblock Class noise vs. attribute noise: A quantitative study.
\newblock {\em Artificial Intelligence Review}, 22(3):177--210, 2004.

\bibitem{zulkernain2010mobile}
Sina Zulkernain, Praveen Madiraju, Sheikh~Iqbal Ahamed, and Karl Stamm.
\newblock A mobile intelligent interruption management system.
\newblock {\em J. UCS}, 16(15):2060--2080, 2010.

\end{thebibliography}

\end{document}